\documentclass[aps,prb,twocolumn,superscriptaddress]{revtex4-2}
\usepackage{graphics,graphicx, array, afterpage}
\usepackage{qcircuit,amsmath}
\usepackage{grffile}
\usepackage[export]{adjustbox}
\usepackage{lineno}
\usepackage{xcolor}
\usepackage{longtable}
\usepackage{braket}
\usepackage{array}
\usepackage{multirow}
\usepackage{enumerate}
\usepackage{amsmath}
\usepackage{amssymb}
\usepackage{amsthm}
\usepackage{times}
\usepackage{latexsym}
\usepackage{multirow}

\usepackage[colorlinks, citecolor=blue]{hyperref}
\hypersetup{linkcolor=magenta,urlcolor=blue,citecolor=blue,pdfstartview={FitH},urlcolor=blue}

\begin{document}

\title{Exact Quantum Many-Body Scars in Higher-Spin Kinetically Constrained Models}

\author{Dong Yuan}
\thanks{These authors contributed equally to this work.}
\affiliation{Center for Quantum Information, IIIS, Tsinghua University, Beijing 100084, People's Republic of China}

\author{Shun-Yao Zhang}
\thanks{These authors contributed equally to this work.}
\affiliation{Center for Quantum Information, IIIS, Tsinghua University, Beijing 100084, People's Republic of China}

\author{Dong-Ling Deng}
\email{dldeng@tsinghua.edu.cn}
\affiliation{Center for Quantum Information, IIIS, Tsinghua University, Beijing 100084, People's Republic of China}
\affiliation{Shanghai Qi Zhi Institute, 41st Floor, AI Tower, No. 701 Yunjin Road, Xuhui District, Shanghai 200232, China}
\affiliation{Hefei National Laboratory, Hefei 230088, People's Republic of China}

\begin{abstract}
We discover a variety of exact quantum many-body scars in higher-spin kinetically constrained models, through the recently developed DMRG-S algorithm [Zhang \emph{et al.}, Phys. Rev. Lett. 131, 020402]. Specifically, for the higher-spin PXP model on arbitrary bipartite lattices of any spatial dimension, we find exact many-body scars that are equidistantly spaced in the energy spectrum and exhibit similar structures to the ground state of the Affleck-Kennedy-Lieb-Tasaki model. For the one-dimensional Fermi-Hubbard model with a tilted potential in a certain parameter regime, whose effective model is equivalent to a kinetically constrained spin model with four degrees of freedom on each site, we find several many-body scars at energy $E=0$ and $E=\pm \sqrt{2}$ that can be exactly represented as matrix product states with finite bond dimensions. Our results demonstrate that larger local degrees of freedom in the kinetically constrained models provide a much broader space for the emergence of quantum many-body scars and weak ergodicity breaking.


\end{abstract}

\maketitle 
\section{Introduction}
In isolated quantum many-body systems, long-time evolution governed by 
non-integrable Hamiltonians typically causes local observables to approach their thermal expectation values. This kind of quantum thermalization dynamics can be illustrated by the eigenstate thermalization hypothesis (ETH)~\cite{Deutsch1991Quantum,Srednicki1994Chaos,Rigol2008Thermalization,Deutsch2018Eigenstate}: The subsystem reduced density matrices of typical excited eigenstates are close to the Gibbs ensembles at the temperature set by the eigenenergy. 
Known strong violation of the ETH paradigm includes the integrable~\cite{Sutherland2004beautiful} and many-body localized~\cite{Nandkishore2015Manybody,Abanin2019Colloquium} systems, in which either exact or approximate extensive conserved quantities prevent the systems from thermalization.
Recently, experiments in Rydberg-atom
quantum simulators demonstrated unexpected long-time coherent revival dynamics from certain special initial states~\cite{Bernien2017Probing,Bluvstein2021Controlling}. This type of weak ergodicity breaking has been attributed to a small fraction of non-thermal excited eigenstates immersed in a sea of thermal ones, dubbed quantum many-body scars~\cite{Turner2018weak,Turner2018quantum,Moudgalya2021Quantum,Chandran2023Quantum}. 

Quantum many-body scarred eigenstates with exact analytical expressions have been found and constructed in various models~\cite{Shiraishi2017Systematic,Moudgalya2018Exact,Moudgalya2018Entanglement,Choi2019emergent,Schecter2019Weak,lin2019exact,ok2019topological,Iadecola2020Quantum,Chattopadhyay2020quantum,Moudgalya2020Large,Lin2020Quantum,Pakrouski2020Many,Pakrouski2021Group,Ren2021Quasisymmetry,Lee2020Exact,Chertkov2021Motif,Surace2020Weak,Surace2021Exact,karle2021area,Langlett2021Hilbert,Langlett2022rainbow,Banerjee2021Quantum,Biswas2022Scars,Schindler2022Exact}. In the previous work~\cite{Zhang2023Extracting}, by leveraging the sub-volume-law entanglement entropy of many-body scars, we proposed the DMRG-S algorithm to systematically obtain accurate matrix product state (MPS) representations for scars in generic Hamiltonians without \textit{a priori} knowledge. Here, we present a variety of exact quantum many-body scars found by DMRG-S in higher-spin kinetically constrained models. Previous works (Ref.~\cite{Ho2019periodic} and~\cite{Desaules2021Proposal}) numerically showed that these models host scarred eigenstates (without exact analytical expressions) and exhibit revival dynamics from special initial states, yet overlooked the existence of these exact scars. The larger Hilbert space dimension and eigensubspace degeneracy~\cite{Schecter2018Many,karle2021area,buijsman2022number} of higher-spin models pose notorious numerical challenges for finding highly excited scarred eigenstates through exact diagonalization, which are exactly overcome by our DMRG-S algorithm. Our results demonstrate that the larger local degrees of freedom in kinetically constrained models~\cite{Garrahan2010Kinetically} can provide a much broader space for the emergence of quantum many-body scars and inducing weak ergodicity breaking in these models.

In Sec.~\ref{sec:PXP} for the higher-spin PXP models~\cite{Ho2019periodic,Zhang2023Extracting} on arbitrary bipartite lattices of any spatial dimension, we find exact many-body scars that are equidistantly spaced in the energy spectrum (i.e., forming a tower of scarred eigenstates), which exhibit simple structures similar to the ground state of the Affleck-Kennedy-Lieb-Tasaki (AKLT) model \cite{Affleck1987Rigorous}. 
We further write down related Hamiltonians on arbitrary lattices that host these AKLT-like scars with area-law entanglement.
In Sec.~\ref{sec:tilted_FH}, we re-examine the one-dimensional (1D) Fermi-Hubbard model with a tilted potential in the parameter regime $U = \Delta \gg J$ and the filling factor $\nu=1$~\cite{Desaules2021Proposal}, whose effective model is equivalent to a kinetically constrained spin model with four degrees of freedom on each site. We find several many-body scars at energy $E=0$ and $E=\pm \sqrt{2}$ that have exact MPS representations with finite bond dimensions (similar to those found in the 1D spin-$1/2$ PXP model~\cite{lin2019exact}). After being projected into the corresponding symmetry sectors, these scarred eigenstates possess logarithmic entanglement entropy scaling.
We provide the concluding remarks and outlooks in Sec.~\ref{sec:conclusion}. More details about the DMRG-S algorithm and property analyses of the scars are presented in the Appendices.

\section{Exact many-body scar towers in higher-spin PXP models}
\label{sec:PXP}

\begin{figure}
\centering
\includegraphics[width=1.0\linewidth]{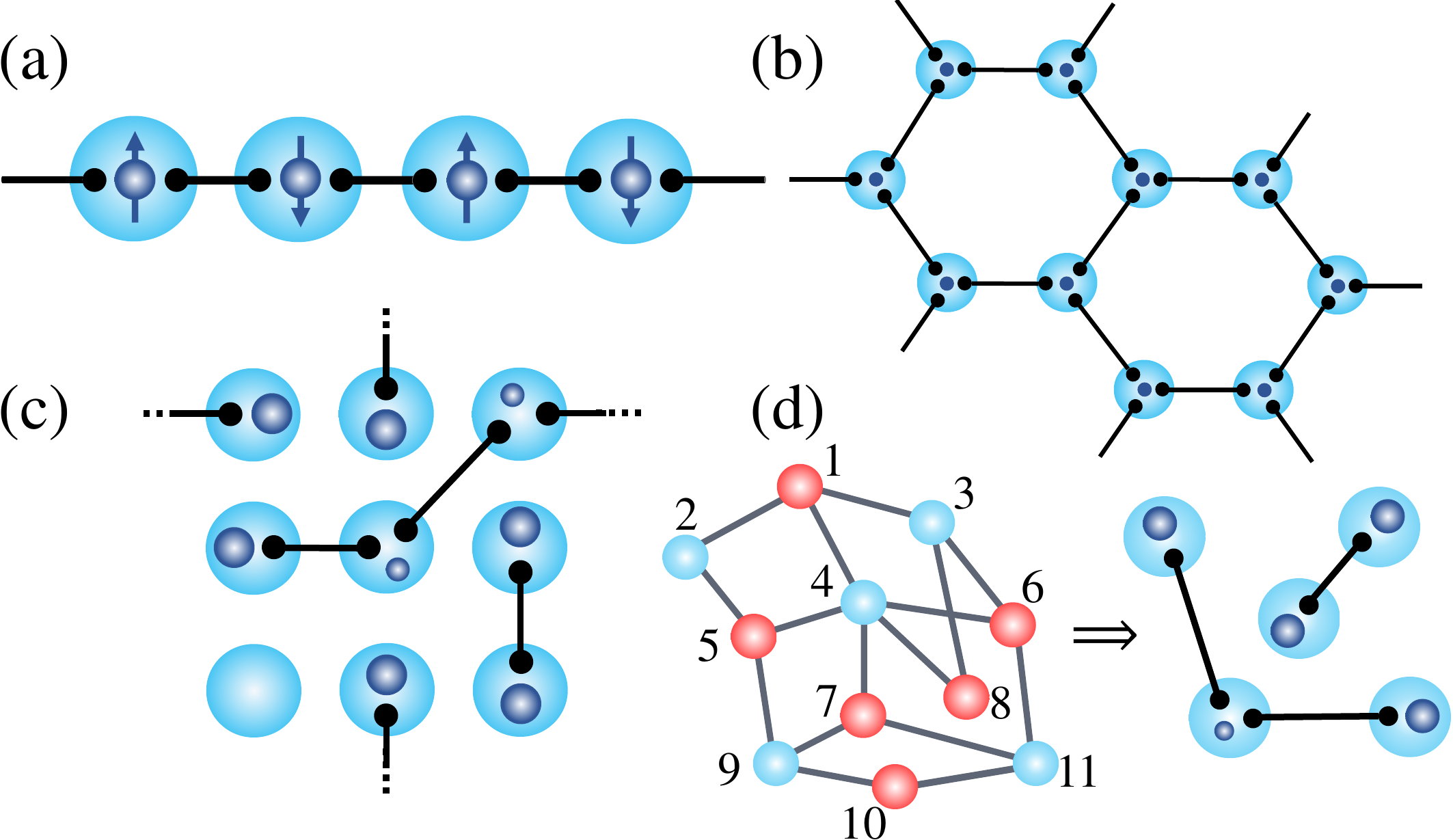} 
\caption{A schematic illustration for the exact quantum many-body scars in the higher-spin PXP models on generic bipartite lattices. The black circles denote the spin-$1/2$'s split out to form singlets (black lines).
The dark blue circles denote the freely-rotating smaller spins. In (a)-(c) we only display the structure of states on one sublattice (marked by light blue). The state on the other sublattice (marked by red) is the direct product of states projected by the $\{P_i\}$ operators.
(a) The AKLT-like scar states on the one-dimensional spin chain. (b) The AKLT-like scar states on the two-dimensional honeycomb lattice with the Lieb-lattice-like structure (i.e. the other set of vertices are located on the edges of the honeycomb lattice). (c) Possible dimer covering configurations for scar states on the two-dimensional square lattice. We take the $3\times 3$ periodic boundary sublattice as an example. (d) On irregular bipartite graphs, the minimum dimer covering configurations on one sublattice are determined by the connection relationship with the other sublattice. 
}
\label{fig:PXP_illustra}
\end{figure}

In this section, we consider the following spin-$s$ PXP Hamiltonian on a generic bipartite graph $G = (V, E)$, where $V$ is the set of all the vertices (divided into two sets $A$ and $B$) and $E$ is the set of all the edges (only including interconnections between $A$ and $B$).
\begin{equation}
H_\text{PXP} = \sum_{i\in V} S^x_i \left(\prod_{(i, j) \in E} P_j\right)
\end{equation}
The spin-$s$ on each site $i$ contains $2s+1$ bases $\{\ket{s, m^z = -s},\ket{s, m^z = -s+1},\cdots,\ket{s, m^z = s-1},$ $\ket{s, m^z = s}\}$. $\ket{s, m^z}$ denotes the eigenstate with the total spin equal to $s$ and the $z$-direction polarization equal to $m^z$. $P_i=\ket{s, m^z = -s}_i\bra{s, m^z = -s}$.
$S^x$ is the $x$-direction angular momentum operator of spin-$s$.
$\bra{s,m^z \pm 1}S^x\ket{s,m^z}=\sqrt{(s\pm m^z +1)(s \mp m^z)}/2$. 

In particular, the spin-$1/2$ PXP Hamiltonians on one~\cite{Turner2018weak} and two~\cite{Lin2020Quantum} dimensional lattices were proposed to describe the Rydberg atom arrays in the nearest-neighbor blockade regime~\cite{Bernien2017Probing,Bluvstein2021Controlling} (i.e., neighboring atoms can not both be in the Rydberg excited states). Previous works generalized the PXP Hamiltonians to higher spins~\cite{Ho2019periodic,Desaules2022Weak,Desaules2022Prominent}, which could possibly be realized through multiple atomic internal states. These works numerically found towers of scarred eigenstates (without exact analytical expressions) and revealed their corresponding oscillatory dynamics from special initial states.
Besides, in our previous work~\cite{Zhang2023Extracting} through the DMRG-S algorithm, we found one exact many-body scar with a bond-dimension-two MPS representation in the $E=0$ nullspace of the 1D PXP models of integer spin-$s$. 
Here, we further show that general spin-$s$ PXP Hamiltonians on any bipartite graph of any spatial dimension host exact many-body scars forming an equally spaced tower in the energy spectrum.

We begin by considering the spin-$s$ PXP Hamiltonians on the 1D spin chain with the periodic boundary condition, which has been shown to be chaotic and nonintegrable by level statistics calculations~\cite{Ho2019periodic}.
\begin{equation}
H_\text{PXP} = \sum_{i=1}^L P_{i-1} S^x_i P_{i+1},
\label{Eq:PXP_Ham_1D}
\end{equation}
where $L$ is the number of spin-$s$'s (we consider the case of even $L$). We show that the following states are exact quantum many-body scarred eigenstates, with the pictorial illustration shown in Fig.~\ref{fig:PXP_illustra}(a).
\begin{equation}
\ket{\Psi} = \ket{\Psi_\text{odd}} \otimes \left(\bigotimes_{ i \in \text{even} } \ket{s, m^z=-s}_i \right) 
\end{equation}
\begin{equation}
\ket{\Psi_\text{odd}} = \left( \prod_{ i \in \text{odd}} \Pi_i^{s} \right) \left(\bigotimes_{ i \in \text{odd} } \ket{ s-1, m^x_i }_i \bigotimes_{ i \in \text{odd} } \ket{\chi}_{i,i+2} \right).
\end{equation}
$\ket{\chi}_{i,j} = (\ket{\uparrow \downarrow} - \ket{\downarrow \uparrow})_{i,j}/\sqrt{2}$ denotes the spin singlet state (dimer) formed by two spin-$1/2$'s on the sites $i,j$.
$ \Pi_i^s $ projects the two spin-$1/2$'s and one spin-$(s-1)$ on the site $i$ to a total spin-$s$. We can exchange the role of even and odd sites to obtain another set of exact scars.

To prove that all the $\ket{\Psi}$'s are eigenstates of the Hamiltonian Eq.~\eqref{Eq:PXP_Ham_1D}, we observe that
\begin{equation}
\begin{aligned}
\left( \sum_{ i \in \text{odd} } P_{i-1} S^x_i P_{i+1} \right) \ket{\Psi} 
& = \left( \sum_{ i \in \text{odd} }  S^x_i  \right) \ket{\Psi} \\
& = \left( \sum_{i \in \text{odd} } m^x_i \right) \ket{\Psi},
\end{aligned}
\label{Eq:even_PXP_sum}
\end{equation}
since the projector $\left( \prod_{ i \in \text{odd}} \Pi_i^s \right)$ preserves the total $x$-direction angular momentum before and after the projection. 

We also have the relations
\begin{equation}
P_i P_{i + 2} \ket{\Psi_\text{odd}} = 0\quad \forall \ \text{odd } i.
\label{Eq:odd_PP_annihilate}
\end{equation}
Notice that $P_i P_{i + 2}$ projects the two spin-$s$'s on the sites $i,i+2$ onto a total spin-$2s$ with $m^z = -2s$. However, since we have already split two spin-$1/2$'s from the two spin-$s$'s to form a singlet state $\ket{\chi}_{i,i+2}$, $P_i P_{i + 2}$ simply annihilates $ \ket{\Psi_\text{odd}} $. The constructions of these exact scars retain the same spirit as the ground state of the AKLT model, despite the fact that here we further add a freely-rotating spin-$(s-1)$ on each site.

Combining Eq.~\eqref{Eq:even_PXP_sum} and Eq.~\eqref{Eq:odd_PP_annihilate} together, we prove that $\ket{\Psi}$ is the eigenstate of the 1D spin-$s$ PXP Hamiltonian with eigenenergy $E = \sum_{i \in \text{odd} } m^x_i$, where $m^x_i$ can take values $\{-(s-1),-(s-2),\cdots,s-2, s-1 \}$. The bipartite entanglement entropy of $\ket{\Psi}$ (or $\ket{\Psi_\text{odd}}$) equals $\ln 2$ given by the singlet states. All these $\ket{\Psi}$'s can be exactly represented as MPSs with bond dimensions $\chi=2$. We provide the MPS expressions for the $\ket{\Psi}$ states of the 1D spin-$1$ and spin-$3/2$ PXP model in Appendix.~\ref{sec:MPS_3_2_PXP}.
We then deduce that these $(2s-1)^L$ AKLT-like states with area-law entanglement form an equidistantly spaced many-body scar tower in the energy spectrum. A natural consequence reflecting on the quench dynamics is that if we start the PXP Hamiltonian evolution from initial states like $\left( \prod_{ i \in \text{odd}} \Pi_i^{s} \right) \left(\bigotimes_{ i \in \text{odd} } \ket{ s-1, m^z_i }_i \bigotimes_{ i \in \text{odd} } \ket{\chi}_{i,i+2} \right) \otimes \left(\bigotimes_{ i \in \text{even} } \ket{s, m^z=-s}_i \right) $ [all the freely-rotating spin-$(s-1)$'s take eigenstates of the $z$-direction angular momentum], we will obtain perfect periodic oscillations.

We can directly generalize the above constructions to the PXP Hamiltonians on any bipartite graph of any spatial dimension, as shown in Fig.~\ref{fig:PXP_illustra}(b)-(d). In particular, we first consider a special kind of bipartite lattices similar to the Lieb lattice~\cite{Weeks2010Topological}: Imagine that the vertices of the sublattice $A$ constitute a graph $G_A = (V_A, E_A)$. The vertices of the sublattice $B$ are put on the midpoints of all the edges in $E_A$.
For example, Fig.~\ref{fig:PXP_illustra}(b) shows the honeycomb lattice formed by vertices in the sublattice $A$ (marked by light blue), and the vertices of the sublattice $B$ (marked by red, omitted in this subfigure) are located on all the edges of the honeycomb lattice. For PXP Hamiltonians on this type of bipartite graphs, we could construct the exact many-body scars by fixing all the spin-$s$'s on the sublattice $B$ to be $\ket{s, m^z = -s}$.
The constraints for the states on the sublattice $A$ then become 
\begin{equation}
P_i P_j \ket{\Psi_A} = 0, \quad \forall \  (i,j) \in E_A.
\label{Eq:PP_constraint}
\end{equation}
We thus need to put one singlet $\ket{\chi}_{i,j}$ on all the edges in $E_A$ to fulfill these requirements. Finally we obtain the AKLT-like dimer covering structure for the spin-$s$'s on the sublattice $A$, as illustrated in Fig.~\ref{fig:PXP_illustra}(b).

Inspired by the derivations above, we can further ignore the sublattice $B$ and write down the following spin-$s$ ``XPP"-type Hamiltonian on a generic graph $G=(V,E)$:
\begin{equation}
H_\text{XPP}= \sum_{i \in V} S^x_{i} + \sum_{(i,j) \in E} V_{i,j} P_{i} P_{j},
\label{Eq:XPP_Ham}
\end{equation}
where $V_{i,j} = P_i P_j V_{i,j}'$ and $V_{i,j}'$ is a generic two-spin Hermitian operator acting on the sites $i,j$ (we can also replace $V_{i,j}$ with generic two-spin Hermitian operators $V_{i',j'}$ acting on other neighboring sites $i',j'$). One can straightforwardly follow the proofs above to deduce that the XPP Hamiltonian hosts the following AKLT-like quantum many-body scars
\begin{equation}
\ket{\Psi} = \left( \prod_{ i \in V } \Pi_i^{s} \right) \left(\bigotimes_{ i \in V } \ket{ s - z_i / 2, m^x_i }_i \bigotimes_{ (i,j) \in E} \ket{\chi}_{i,j} \right),
\end{equation}
where $z_i$ is the coordinate number of the site $i$ (how many edges connect to the vertex $i$).
Other eigenstates of the XPP Hamiltonian which are not annihilated by all the $P_i P_j$ operators will be affected by the random operators $V_{i,j}$ and get thermalized.
Note that while the XPP Hamiltonian looks similar to the toy model proposed in Ref.~\cite{Choi2019emergent}, it differs from that model intrinsically because the XPP Hamiltonian is beyond the Shiraishi-Mori embedding formalism~\cite{Shiraishi2017Systematic} by $[\sum_{i} S^x_{i}, P_{i} P_{j} ] \neq 0$. Another utility of the XPP Hamiltonian is that, by the random operators $V_{i,j}$ it addresses the potential caveat whether the higher-spin PXP models on generic bipartite graphs are indeed chaotic and nonintegrable. Previous level statistics calculations for small $s$ in 1D~\cite{Ho2019periodic} and 2D~\cite{Lin2020Quantum} have given affirmative answers. However, conducting similar calculations for larger $s$ and generic graphs is numerically challenging due to the even larger Hilbert space dimension.


Now we consider the PXP Hamiltonians on more generic higher-dimensional bipartite lattices, where the constraints for the states on the sublattice $A$ are not always two-body like Eq.~\eqref{Eq:PP_constraint}.
As an example for the 2D square lattice, if we fix all the spin-$s$'s on the sublattice $B$ to be $\ket{s, m^z = -s}$, the constraints for the states on the sublattice $A$ become $\left(\prod_{ i \in \Box} P_i \right)  \ket{\Psi_A} = 0$, where $i \in \Box$ denotes the four vertices in one square plaquette of the sublattice $A$. Hence, we need to put one dimer on each plaquette, possibly on the one of the four edges or one of the two diagonals, to fulfill the constraints. One illustrative dimer covering configuration for the scar states are displayed in Fig.~\ref{fig:PXP_illustra}(c), where we take the $3 \times 3$ periodic boundary sublattice as an example. 
In order to construct the scarred eigenstates as many as possible, we should put dimers on the sublattice as few as possible. 
Moreover, the scarred eigenstates with more than one dimer on certain plaquettes can be obtained through appropriate superposition of the scarred eigenstates with only one dimer on each plaquette (further splitting out spin-$1/2$'s from the freely-rotating spins to form singlets).
Note that the dimers put on the edges of plaquettes can be shared by two neighboring plaquettes. 

For PXP models on irregular bipartite graphs as shown in Fig.~\ref{fig:PXP_illustra}(d), we provide several basic rules for constructing the minimum dimer covering configurations on the sublattice $A$ (marked by light blue): When fixing all the spin-$s$'s on the sublattice $B$ (marked by red) to be $\ket{s, m^z = -s}$ and all the freely-rotating smaller spins (dark blue circles) to point to the $x$-direction, the PXP Hamiltonian reduces to the constraints imposed on the spin-$s$'s of the sublattice $A$. We start to form the singlet states $\ket{\chi}_{i,j}$ induced by the $B$ vertices with the fewest edges. For instance, the vertex $8$ and vertex $10$ in Fig.~\ref{fig:PXP_illustra}(d) lead to the dimers $\ket{\chi}_{3,4}$ and $\ket{\chi}_{9,11}$. Next we check whether the constraints induced by other $B$ vertices have already been satisfied, e.g., the constraints induced by the vertices $1,6,7$. We exclude the redundant $B$ vertices and iterate to the next $B$ vertices with the fewest edges. Note that if there exist some dangling $B$ vertices (only connecting to one $A$ vertex), a dimer should be created inside that $A$ vertex.

Several remarks come in order. First, we mention that the existence of these AKLT-like scars in the PXP models requires that the spin-$s$ should be larger than certain values depending on the connection relationship of the underlying graphs. For example, for the PXP Hamiltonian on the 1D spin chain, $s$ should be equal or larger than $1$. For the 2D honeycomb lattice with the Lieb-lattice-like structure, $s$ should equal or larger than $3/2$. It demonstrates that larger local degrees of freedom provide a much broader space for the emergence of quantum many-body scars. 
Second, we notice that the AKLT-like scars with the same eigenenergy are not orthogonal to each other due to the projections by the $\{ \Pi^s_i \}$ operators. However, for a fixed dimer covering configuration $\{ \ket{\chi}_{i,j} \}$, different $\ket{\Psi}$ taking different $\{ m_i^x \}$ values are linear independent, because the projections by $\{ \Pi^s_i \}$ are onsite and the underlying freely-rotating spins have gone through all the eigenstates of the $x$-direction angular momentum (also numerically verified). Third, these AKLT-like scars will still appear if we globally rotate the $S^x$ operators in the PXP Hamiltonian, i.e., $ R^{\dagger} S^x R $, $R = \exp( - i \theta \hat{n} \cdot \hat{S} ) $, $\hat{n}=(n^x,n^y,n^z)$ and $\hat{S}=(S^x,S^y,S^z)$. We only require the freely-rotating spins to take the eigenstates of the corresponding $\hat{n}$-direction angular momentum operators.
Finally, the singlet (dimer) covering structures of the above scarred eigenstates are reminiscent of the constructions of product-state scars in other kinetically constrained models~\cite{Lin2020Quantum,Surace2021Exact,Surace2020Weak}, yet here we focus on the higher-spin cases and these AKLT-like scars possess non-zero entanglement entropy.


\section{Exact many-body scars in the 1D tilted Fermi-Hubbard model}
\label{sec:tilted_FH}

In this section, we consider the following 1D Fermi-Hubbard model with a tilted potential, which
has been experimentally realized by ultracold atoms in the optical lattice~\cite{Scherg2021Observing}. The model exhibits scarred revival dynamics in the parameter regime $U\approx \Delta\gg J$ with the electronic filling factor $\nu=1$ \cite{Desaules2021Proposal}:
\begin{equation}
\begin{aligned}
H = \sum_{j, \sigma=\uparrow,\downarrow} &  ( -J c^{\dagger}_{j,\sigma} c_{j+1,\sigma} + h.c. \\
&+ \Delta j n_{j,\sigma} ) + U \sum_j n_{j,\uparrow} n_{j,\downarrow}.     
\end{aligned}  
\end{equation}
The above Hamiltonian conserves the number of spin-up and spin-down fermions. We particular consider the case of $L/2$ spin-up fermions and $L/2$ spin-down fermions hopping on the 1D lattice with $L$ sites.
When $U \approx \Delta\gg J$ the Hilbert space fragments into several dynamically disconnected subspaces labelled by the sum of the number of doublons ($U$ terms, $\ket{\updownarrow}_j$) and the dipole moment ($\Delta$ terms, $j n_j$). Below we consider the case of $J=1$ and $U = \Delta$.

\begin{figure}
\centering
\includegraphics[width=.9\linewidth]{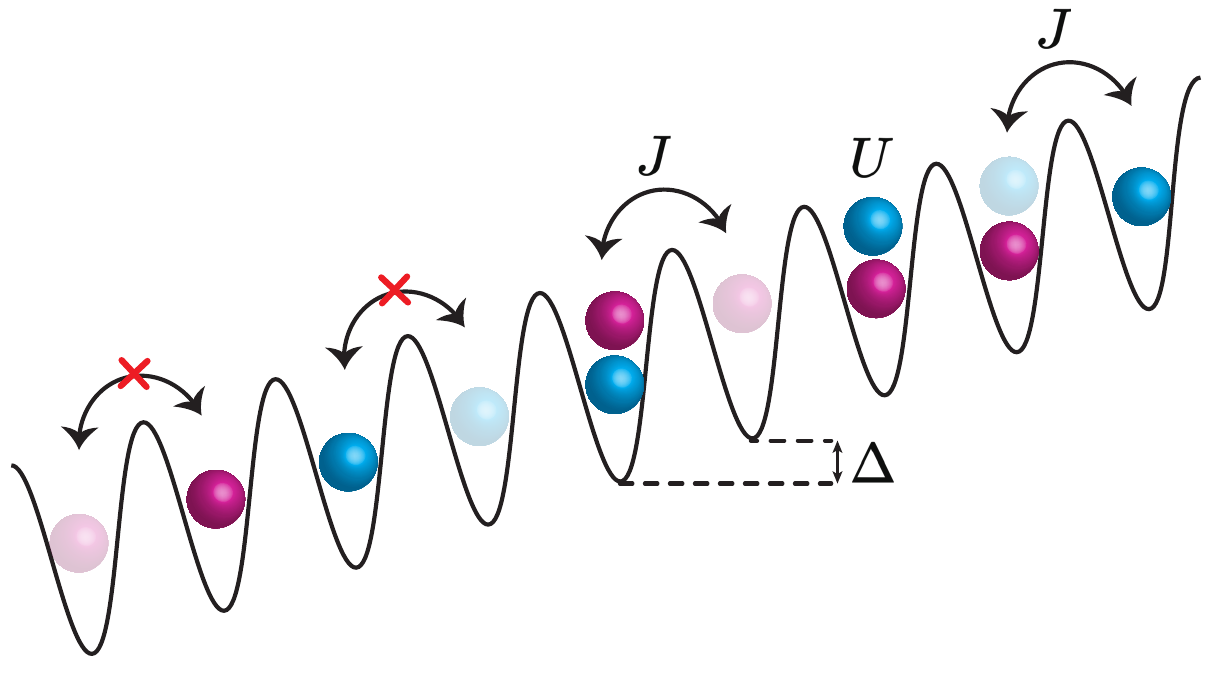} 
\caption{An illustration of the constrained hopping in the 1D tilted Fermi-Hubbard model in the parameter regime $U\approx \Delta \gg J$. The left (right) hopping of spinful fermions is only allowed if the hopping creates (breaks) a doublon.}
\label{fig:Tilted}
\end{figure}

In each fragmented sector, the hopping terms can be treated perturbatively by the Schrieffer-Wolff transformation~\cite{Bravyi2011Schrieffer}. To the leading order, the allowed hopping Hamiltonian becomes
\begin{equation}
H_{\mathrm{eff}}=-\sum_{j, \sigma=\uparrow,\downarrow} \left[c_{j,\sigma}^\dagger c_{j+1, \sigma} n_{j, \bar{\sigma}} (1 - n_{j+1, \bar{\sigma}}) + h.c.\right],
\label{Eq:H_eff_fermion}
\end{equation}
where $n_{j,\sigma} = c_{j,\sigma}^\dagger c_{j, \sigma}$, and $\bar{\sigma}$ denotes the opposite spin of $\sigma$. The effective Hamiltonian simply reads that the left (right) hopping is only allowed if the hopping creates (breaks) a doublon (see Fig.~\ref{fig:Tilted} for a pictorial illustration), as required by the conservation of $U + \Delta$ terms. 

In order to apply the DMRG-S algorithm to find exact scars, we carry out the Jordan-Wigner transformation to rewrite the effective Hamiltonian Eq.~\eqref{Eq:H_eff_fermion} on the spin bases. 
Note that since we are dealing with spinful fermions, we need to conduct the following Jordan-Wigner transformation for two sets of fermions~\cite{Desaules2021Proposal}:
\begin{equation}
\begin{aligned}
c_{ j,\uparrow} & = \prod_{i<j}\left( S^z_{i,\uparrow} S^z_{i,\downarrow} \right) S_{j,\uparrow }^- \\ 
c_{ j,\downarrow} & = \prod_{i<j}\left( S^z_{i,\uparrow} S^z_{i,\downarrow} \right) (-S^z_{j,\uparrow}) S_{j,\downarrow }^- \\
n_{j, \sigma} & = c_{ j,\sigma }^\dagger c_{ j,\sigma } = \frac{ 1+ S^z_{j,\sigma} }{2} := P_{j,\sigma}.
\end{aligned}
\end{equation}
Here we use the excited states in spin bases to represent the occupied states in fermion bases. $S^\alpha \  (\alpha = x,y,z)$ are spin-$1/2$ angular momentum operators and $S^{\pm} 
= S^x \pm i S^y$ are the raising and lowering operator.
The effective Hamiltonian then becomes a kinetically constrained spin model with four degrees of freedom
on each site (the vacuum state $|0\rangle$, the spin-up state $\ket{\uparrow}$,  the spin-down state $\ket{\downarrow}$ and the doubly occupied state $|\updownarrow\rangle$):
\begin{equation}
\begin{aligned}
H_{\mathrm{eff}} &= \sum_{j} \left[S_{j,\uparrow}^+ S_{j+1, \uparrow}^- P_{j, \downarrow} (1 - P_{j+1, \downarrow}) + h.c.\right] \\
& - \sum_{j} \left[ S_{j,\downarrow}^+ S_{j+1, \downarrow}^- P_{j, \uparrow} (1 - P_{j+1, \uparrow}) + h.c.\right].   
\end{aligned}
\label{Eq:H_eff_spin}
\end{equation}




Ref.~\cite{Desaules2021Proposal} numerically showed that the above effective Hamiltonian hosts towers of quantum many-body scars (not providing exact analytical expressions) which are responsible for the revival dynamics from special initial states. We employ the DMRG-S method to search across the whole spectrum, and discover the following scarred eigenstates at the energy $E=0$ and $E=\pm \sqrt{2}$ with exact MPS representations: 
\begin{equation}
\left|\Lambda_{\alpha, \beta}\right\rangle=\sum_{\{\mu_i\}} v_\alpha B^{\mu_1} C^{\mu_2} \cdots B^{\mu_{L-1}} C^{\mu_L} v_\beta^T 
 \left|\mu_1 \mu_2 \cdots \mu_L\right\rangle
\label{Eq:24_42_scar}
\end{equation}
where $\alpha, \beta \in \{+, -\}$,
$v_+=(1,1)$, $v_-=(1,-1)$ and 
each $\mu_i$ could take the four possible bases $\{0,\uparrow,\downarrow,\updownarrow\}$,
\begin{equation}
\begin{aligned}
& B^{[0]} = \left(\begin{array}{llll}
0 & 0 & 0 & 0 \\
0 & 0 & 0 & 0
\end{array}\right), 
\quad  
C^{[0]}=\frac{1}{\sqrt{2}}\left(\begin{array}{llll}
0 & 0  & 1 & 1\\
0 & 0 & 1 & -1 
\end{array}\right)^T,\\
& B^{[\uparrow]} = \left(\begin{array}{llll}
0 & -1 & 0 & 0 \\
0 & 0 & 0 & 0
\end{array}\right),
\quad 
C^{[\uparrow]}=\left(\begin{array}{llll}
\sqrt{2} & 0 & 0 & 0\\
0 & 0 & 0 & 0
\end{array}\right)^T,\\
& B^{[\downarrow]} = \left(\begin{array}{llll}
0 & 0 & 0 & 0 \\
1 & 0 & 0 & 0
\end{array}\right),
\quad 
C^{[\downarrow]}=\left(\begin{array}{llll}
0 & 0 & 0 & 0\\
0 & -\sqrt{2} & 0 & 0
\end{array}\right)^T,\\
& B^{[\updownarrow]} = \frac{1}{\sqrt{2}}\left(\begin{array}{llll}
0 & 0 & 1 & 1 \\
0 & 0 & -1 & 1
\end{array}\right),
\quad 
C^{[\updownarrow]}=\left(\begin{array}{llll}
0 & 0 & 0 & 0\\
0 & 0 & 0 & 0
\end{array}\right)^T.
\end{aligned}
\label{Eq:B_C_matrix}
\end{equation}

Below we rigorously prove that $H_{\mathrm{eff}} \left|\Lambda_{\pm, \pm}\right\rangle = 0$,
$H_{\mathrm{eff}} \left|\Lambda_{+, -}\right\rangle =\sqrt{2} \left|\Lambda_{+, -}\right\rangle$ and $H_{\mathrm{eff}} \left|\Lambda_{-, +}\right\rangle = -\sqrt{2} \left|\Lambda_{-, +}\right\rangle$. 
We notice that these scarred eigenstates share similar structures with the exact scars found in the 1D spin-$1/2$ PXP model~\cite{lin2019exact}, which we further demonstrate in the proofs below.

Similar to the techniques used in Ref.~\cite{lin2019exact}, we group the neighboring two sites $[2b-1, 2b]$ into a block $b$ $(b=1,2,\cdots,L/2)$. From Eq.~\eqref{Eq:B_C_matrix} we find that the block states with non-zero matrices $A^{[2b-1, 2b]}$ of each block are $\ket{\uparrow,\downarrow},\ket{\updownarrow,0},\ket{\downarrow,\uparrow}$:
\begin{equation}
A^{[\uparrow,\downarrow]}  = \sqrt{2}\sigma^+, \quad  A^{[\updownarrow,0]} = \sigma^z,\quad 
A^{[\downarrow,\uparrow]} = \sqrt{2}\sigma^-,
\end{equation}
where $\sigma^{\alpha}$ ($\alpha=x,y,z$) are standard $2\times 2$ Pauli matrices, $\sigma^{\pm} = (\sigma^x \pm i \sigma^y )/2$.
The matrices of other block states are all zero, $A^{[m,n]} = B^{[m]} C^{[n]}= 0_{2\times 2}$,  where $[m,n] \notin \{ [\uparrow,\downarrow],[\updownarrow,0],[\downarrow,\uparrow]\}$.

We also represent the effective Hamiltonian Eq.~\eqref{Eq:H_eff_spin} on the block state bases, which includes the single-block terms
\begin{equation}
\sum_{b=1}^{L/2} (|\updownarrow,0\rangle \langle \downarrow,\uparrow| - | \updownarrow,0\rangle \langle \uparrow,\downarrow |+ h.c. )_b ,
\label{Eq:single_block}
\end{equation}
and the interaction terms between two blocks for the non-zero block states
\begin{align}
\sum_{b=1}^{L/2} h_{b,b+1}  & =  \sum_{b=1}^{L/2} \left[  \left(\ket{\uparrow,\updownarrow} \bra{\uparrow,\downarrow}\right)_b \otimes (\ket{0,\downarrow}\bra{\uparrow,\downarrow})_{b+1} \right.\\
& - \left.\left(\ket{\downarrow,\updownarrow} \bra{\downarrow,\uparrow}\right)_b \otimes (\ket{0,\uparrow}\bra{\downarrow,\uparrow})_{b+1}  + h.c. \right]. \nonumber
\end{align}

Since $(A^{[\uparrow,\downarrow]})^2 = (A^{[\downarrow, \uparrow]})^2 =0 $, we can derive that $h_{b,b+1} \ket{\Lambda_{\alpha, \beta}}=0$, $\forall \ b$.
We then only consider the action of single-block terms on $\left|\Lambda_{\alpha,  \beta}\right\rangle$:
\begin{equation}
\begin{aligned}
H_\text{eff} \left|\Lambda_{\alpha, \beta}\right\rangle = \sum_{b=1}^{L/2} \sum_{\{ d_i \}} ( & v_{\alpha} A^{d_1}  \cdots F^{d_b} \\
& \cdots A^{d_{\frac{L}{2}}} v_{\beta}^T ) \left|d_1 \cdots d_{\frac{L}{2}}\right\rangle,     
\end{aligned}
\label{Eq:H_action}
\end{equation}
where $\{d_i = [m,n] \}_{i=1}^{L/2}$ denote the block state bases, and $F^{d_b}$ are the transformed tensors by the single-block terms in Eq.~\eqref{Eq:single_block}:
\begin{equation}
F^{[\uparrow,\downarrow]} = -\sigma^z,\quad F^{[\updownarrow,0]} = -i\sqrt{2}\sigma^y, \quad F^{[\downarrow,\uparrow]}= \sigma^z.
\end{equation}

It is also straightforward to verify the following relations for the bulk tensors
\begin{equation}
F^{d_b} = (\sigma^x A^{d_b} - A^{d_b} \sigma^x) / \sqrt{2},
\label{Eq:bulk}
\end{equation}
and for the boundary vectors
\begin{equation}
v_\pm F^{d_1} = (\pm v_\pm A^{d_1}  - v_\pm A^{d_1} \sigma^x) /\sqrt{2} ,
\label{Eq:left}
\end{equation}
\begin{equation}
F^{d_{\frac{L}{2}}} v_\pm ^T =  (\sigma^x A^{d_{\frac{L}{2}}} v_\pm^T  \mp   A^{d_{\frac{L}{2}}} v_\pm^T) /\sqrt{2}.
\label{Eq:right}
\end{equation}

By substituting Eq.~\eqref{Eq:bulk}\eqref{Eq:left}\eqref{Eq:right} into Eq.~\eqref{Eq:H_action} and telescoping summing all the terms, we prove that
\begin{equation}
H_\text{eff} \left|\Lambda_{\alpha, \alpha}\right\rangle =0, \ H_\text{eff} \left|\Lambda_{\pm, \mp}\right\rangle =\pm \sqrt{2} \left|\Lambda_{\pm, \mp}\right\rangle.
\end{equation}

\begin{figure}
\centering
\includegraphics[width=.9\linewidth]{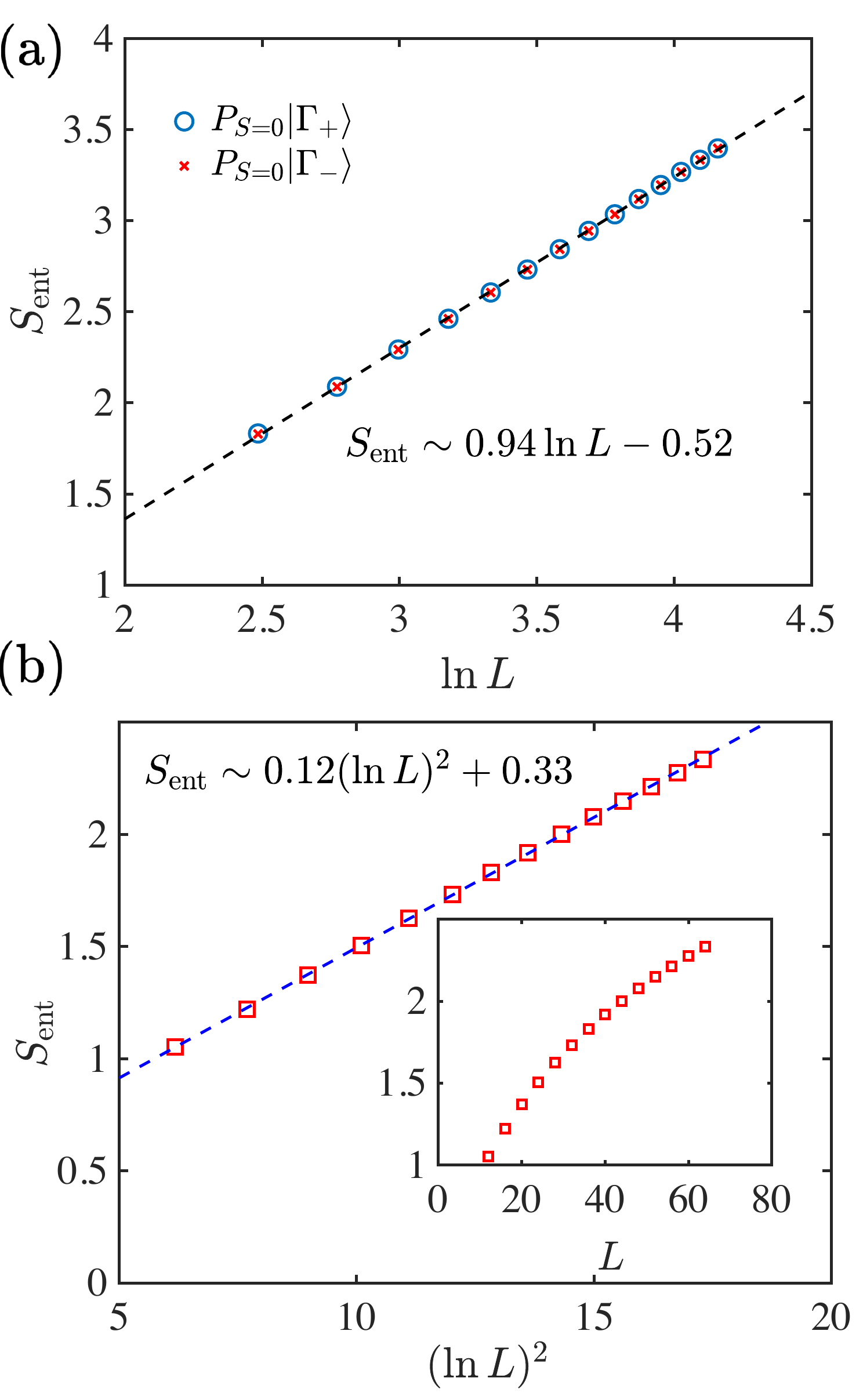} 
\caption{Bipartite entanglement entropy $S_\text{ent}$ scaling of the exact scars in the 1D tilted Fermi-Hubbard model, after being projected into the corresponding symmetry sectors. (a) $S_\text{ent}$ of $P_{S=0} \ket{\Gamma_{\pm}}$ both grow logarithmically with the system size ($L = 12,16,\cdots,64$). (b) $S_\text{ent}$ of $P_{S=1} \ket{\Lambda_{+, -}}$ is instead proportional to 
$(\ln L)^2$ ($L = 12,16,\cdots,64$). The inset displays data in the linear scale.}
\label{fig:scaling}
\end{figure}

As shown in Ref. \cite{Desaules2021Proposal}, 
in the Hilbert subspace dynamically connected to the scar states Eq.~\eqref{Eq:24_42_scar},
the effective Hamiltonian Eq.~\eqref{Eq:H_eff_spin} possesses three symmetries $(\mathbb{Z}_1, \mathbb{Z}_2, \mathbb{S}^2)$, whose concrete expressions are presented in Appendix.~\ref{sec:symm_analysis}. We have verified that $\ket{\Lambda_{\pm, \mp}}$ are eigenstates of $\mathbb{Z}_1$ and $\mathbb{Z}_2$
\begin{equation}
\begin{aligned}
&\mathbb{Z}_1 \left|\Lambda_{\pm, \mp}\right\rangle =- \left|\Lambda_{\pm, \mp}\right\rangle, \\
&\mathbb{Z}_2 \left|\Lambda_{\pm, \mp}\right\rangle =(-1)^{L/2+1} \left|\Lambda_{\pm, \mp}\right\rangle.    
\end{aligned}
\end{equation}
As for $\left|\Lambda_{\pm, \pm}\right\rangle$, the action of $\mathbb{Z}_1$ and $\mathbb{Z}_2$ gives
\begin{equation}
\mathbb{Z}_1 \left|\Lambda_{\pm, \pm}\right\rangle = \left|\Lambda_{\pm, \pm}\right\rangle, 
\   \mathbb{Z}_2 \left|\Lambda_{\pm, \pm}\right\rangle =(-1)^{L/2} \left|\Lambda_{\mp, \mp}\right\rangle.
\end{equation}
We can superpose the states $\left|\Lambda_{+, +}\right\rangle$ and $\left|\Lambda_{-, -}\right\rangle$ to construct the simultaneous eigenstates of $\mathbb{Z}_1$ and $\mathbb{Z}_2$. We define
\begin{equation}
\ket{\Gamma_\pm} = \frac{1}{\sqrt{2}} \left(\ket{\Lambda_{+,+}} \pm \ket{\Lambda_{-,-}} \right),
\end{equation}
which satisfy 
\begin{equation}
\mathbb{Z}_1 \ket{\Gamma_\pm} = \ket{\Gamma_\pm}, 
    \quad \mathbb{Z}_2 \ket{\Gamma_\pm} = \pm (-1)^{L/2} \ket{\Gamma_\pm}.
\end{equation}

Moreover, we define the projectors $ P_S $ to project the above MPSs $\left|\Lambda_{\pm, \mp}\right\rangle$ and $\ket{\Gamma_{\pm}}$ into the symmetry sector $\mathbb{S}^2 = S(S+1)$ of the total angular momentum
\begin{equation}
P_S =  \prod_{j\neq S} \left[\mathbb{S}^2 - j(j+1) \right],
\end{equation}
where $j$ could takes values $0,1,\cdots,L/2$. In Fig.~\ref{fig:scaling} we numerically calculate the  entanglement entropy scaling of these scarred eigenstates within corresponding symmetric sectors. As specific examples, for the $E=0$ scar $\ket{\Gamma_{\pm}}$ with $(\mathbb{Z}_1=+1,\mathbb{Z}_2=\pm 1, \mathbb{S}=0)$, we find that the bipartite entanglement entropy grows logarithmically with the system size $L$ [Fig.~\ref{fig:scaling}(a)]. For the $E=\sqrt{2}$ scar $\ket{\Lambda_{+, -}}$, we project it into the symmetric sector $(\mathbb{Z}_1=-1,\mathbb{Z}_2=-1,\mathbb{S}=1)$ (its even components of $\mathbb{S}=0,2,4,\cdots$ vanish). Interestingly, the entanglement entropy scaling best fits with $S_\text{ent} \sim (\ln L)^2$ [Fig.~\ref{fig:scaling}(b)]. These numerical results demonstrate that the entanglement entropy of theses exact scars scales at most logarithmically with the system size, after being projected into each symmetry sector. 

Finally, if $U \neq \Delta$, the effective Hamiltonian Eq.~\eqref{Eq:H_eff_fermion} additionally acquires a term $(U-\Delta) \sum_j n_{j,\uparrow} n_{j, \downarrow}$, which is reminiscent of the external magnetic field term added to the PXP Hamiltonian~\cite{yao2021quantum}. We have numerically verified that the above MPSs are no longer eigenstates of the new Hamiltonian. We leave the investigations of their stability~\cite{Surace2021Exact} and change of entanglement~\cite{yao2021quantum} to future studies.




\section{Conclusions and Outlooks}
\label{sec:conclusion}
In summary, by utilizing our recently developed DMRG-S algorithm, we numerically searched and discovered several exact quantum many-body scars which were overlooked in two previously studied models. Our results open up a promising avenue towards finding exact scarred eigenstates in kinetically constrained models with larger local degrees of freedom.

Inspired by all the exact scars found in the current paper and Ref.~\cite{Zhang2023Extracting}, one important open question appears that: Whether there always exist or what are the conditions for the existence of certain many-body scars (possibly with simple exact analytical expressions) in other kinetically constrained models with larger local degrees of freedom~\cite{hudomal2020quantum,Desaules2022Weak,Desaules2022Prominent}, especially within their exponentially large degenerate eigensubspaces at $E=0$~\cite{Schecter2018Many,karle2021area,buijsman2022number}. For spin-$1/2$ systems, when assuming the particle-hole symmetry and translation symmetry for two-local Hamiltonians, the question has been affirmatively answered in Ref.~\cite{karle2021area}. However, when extending to higher-spin models, more investigations are needed for future studies.

Besides, the tower of exact AKLT-like scars in the higher-spin PXP models can be represented as bond-dimension-two MPSs, thus are feasible to be prepared on near-term quantum simulators~\cite{Gustafson2023Preparing,Schon2005Sequential,Wei2023Efficient,Murta2023Preparing,Smith2023Deterministic,Chen2023Highfidelity}. Moreover, these scars exist on generic bipartite lattices of any spatial dimension, which could be possibly utilized to yield high-dimensional time-crystalline behaviors under periodic driving~\cite{Bluvstein2021Controlling,Maskara2021discrete,Hudomal2022Driving,Huang2023Engineering,Deng2023Leveraging} (circumventing the possible instability of many-body localization in higher dimensions~\cite{Chandran2016Manybody,De2017Stability,De2017Manybody,Agarwal2017Rare,Potirniche2019Exploration,Gopalakrishnan2019Instability}).

\begin{acknowledgments}
We acknowledge helpful discussions with He-Ran Wang, and previous collaborations with Thomas Iadecola and Shenglong Xu.  
This work was supported by the National Natural Science Foundation of China (Grants No. 12075128 and T2225008) and Shanghai Qi Zhi Institute.
\end{acknowledgments}

\maketitle

\appendix

\section{The DMRG-S algorithm and numerical details}
In Ref.~\cite{Zhang2023Extracting} we developed the DMRG-S algorithm to accurately extract scarred eigenstates by modifying and improving the shift-invert technique~\cite{Luitz2015Manybody,yu2017finding,Serbyn2016Power}.
The intuition for this algorithm is that repeatedly applying the inverse operator $(H-\xi)^{-2}$ to an initial state $\ket{\psi_0}$ eventually yields an eigenstate of $H$ with energy $\xi$. In practice, we represent $\ket{\psi_0}$ as an MPS and consider the sequence of states $\ket{\psi_t}=\mathcal N^{-1}\mathcal A_t^{-1}\ket{\psi_{t-1}}$, where $\mathcal A_t=(H-\xi_t)^{2}$ and $\mathcal N$ is a normalization factor. We require the bond dimension of the state $\ket{\psi_t}$ as $\chi\leq \chi_{\rm max}$, which effectively serves as a filter for states with low entanglement entropy. 
In the iteration step $t$, we circumvent the difficulty of calculating the inverse operator $\mathcal A_t^{-1}$ by variationally optimizing $\ket{\psi_t}$ such that $\braket{\psi_t|\mathcal A_{t}|\psi_t}=\mathcal N^{-1}\braket{\psi_t|\psi_{t-1}}$, where $\mathcal A_t$ can be expressed as a matrix product operator. In the MPS formalism we implement the optimization by locally solving the linear equation
\begin{equation}
     \mathcal{A}_{t,\text{eff}}^{[i,i+1]} \psi_t^{[i,i+1]}=   \tilde{\psi}_{t-1}^{[i,i+1]},
\label{eq:linear}
\end{equation}
where $ \mathcal{A}_{t,\text{eff}}^{[i,i+1]} $ is the local ``effective Hamiltonian" for $\mathcal{A}_t$,  $\psi_t^{[i,i+1]}$ is the local tensor of $|\psi_{t} \rangle$ to be updated, and $\tilde{\psi}_{t-1}^{[i,i+1]}$ is the environment tensor of the overlap  $\langle \psi_t |\psi_{t-1}\rangle$.

We perform the local optimization on each pair of sites $[i,i+1]$ and sweep back and forth. During the iterations, we monitor the energy variance $\sigma_H^2 = \langle H^2 \rangle - \langle H \rangle^2$ of $\ket{\psi_t}$.
We initially set $\xi_0$ within the target energy window $[E-\Delta E, E+\Delta E]$. After a few iterations, if $\sigma_H^2$ reaches a relatively small value (less than $10^{-3}$), we then begin to update $\xi_t=\langle \psi_{t}|H|\psi_{t} \rangle$ during each iteration. Eventually we expect $|\psi_{t} \rangle$  to converge to a scarred eigenstate with the target energy.

In order to extract multiple scarred eigenstates within a degenerate eigensubspace, we adopt the following numerical trick: Suppose we have found the MPS representation of a scar $\ket{\psi_1}$ with the energy $E_1$. We can shift the energy of $\ket{\psi_1}$ by adding one term to the Hamiltonian
\begin{equation}
H \rightarrow H + E_\text{shift} \ket{\psi_1} \bra{\psi_1},
\end{equation}
where $E_\text{shift}$ denotes the shifted energy of $\ket{\psi_1}$. We then apply the DMRG-S algorithm to the new Hamiltonian to extract other scars within the eigensubspace of $E_1$.

For extracting the exact scars in the 1D tilted Fermi-Hubbard model with the filling factor $\nu=1$, we focus on the subspace dynamically connected to the state $| \downarrow \uparrow \uparrow \downarrow, \downarrow \uparrow \uparrow \downarrow, \cdots\rangle$ and its spin-inverted state $| \uparrow \downarrow \downarrow \uparrow, \uparrow \downarrow \downarrow \uparrow, \cdots\rangle$~\cite{Desaules2021Proposal}.
According to the kinetically constrained hopping in Eq.~\eqref{Eq:H_eff_spin}, we find that configurations like $[\updownarrow, \downarrow]_{j,j+1},[\updownarrow, \uparrow]_{j,j+1}, [\updownarrow, \updownarrow]_{j,j+1}, [0, 0]_{j,j+1}, [\downarrow, 0]_{j,j+1}, [\uparrow, 0]_{j,j+1}$ are forbidden to appear in this subspace. Hence, at the end of each DMRG-S iteration we project out all the configurations above, analogous to projecting out the $[\uparrow,\uparrow]_{j,j+1}$ configurations in the 1D spin-$1/2$ PXP model~\cite{Zhang2023Extracting}. We further introduce a large Zeeman-field term $B^z \sum_j (n_{j,\uparrow} - n_{j,\downarrow})$ to the effective Hamiltonian Eq.~\eqref{Eq:H_eff_spin}, which along with excluding the forbidden configuration, is sufficient to maintain the particle conservation of the $L/2$ spin-up fermions and $L/2$ spin-down fermions.

\section{MPS representations of the exact scars in the 1D spin-$1$ and spin-$3/2$ PXP model}
\label{sec:MPS_3_2_PXP}
For the 1D spin-$1$ PXP model, when fixing the states on all the even sites to be $\ket{s = 1, m^z = -1}$, the spin-$1$'s on the odd sites split out two spin-$1/2$'s to form dimers with spin-$1/2$'s on the neighboring odd sites (there are no freely-rotating spins here). That exactly corresponds the ground state of the spin-$1$ AKLT model, which can be represented as the following $\chi=2$ MPS
\begin{equation}
\ket{\Psi_\text{odd}}=\sum_{\{\mu_i\}} \text{Tr} \left[A^{\mu_1} A^{\mu_2} \cdots A^{\mu_L}\right] \ket{\mu_1 \mu_2 \cdots \mu_L},
\end{equation}
\begin{equation}
A^{[1]} = \sqrt{2} \sigma^+, \quad A^{[0]} = - \sigma^z, \quad  A^{[-1]} = - \sqrt{2} \sigma^-.
\label{Eq:AKLT_spin_1}
\end{equation}
$\ket{\Psi_\text{odd}} \otimes \left(\bigotimes_{ i \in \text{even} } \ket{s=1, m^z=-1}_i \right) $ then becomes an $E=0$ scarred eigenstate of 1D spin-$1$ PXP model.

For the AKLT-like scars in larger spin-$s$ PXP models, we could obtain their MPS representations based on the spin-$1$ case. We take the 1D spin-$3/2$ PXP model as an example, while the cases for larger spins and generic graphs could be obtained with similar techniques. 
The only difference compared to the spin-$1$ case is that there remains a freely-rotating spin-$1/2$ on each odd site. In order to construct an eigenstate of the PXP Hamiltonian, we require these freely-rotating spin-$1/2$'s to take $\ket{s=\frac{1}{2}, m^x = \pm \frac{1}{2}} = (\ket{s=\frac{1}{2}, m^z = \frac{1}{2}} \pm \ket{s=\frac{1}{2}, m^z =  - \frac{1}{2}})/\sqrt{2}$. Then we multiply the spin-$1$ matrices Eq.~\eqref{Eq:AKLT_spin_1} with the $\pm \sqrt{1/2}$ in $\ket{s=\frac{1}{2}, m^x = \pm \frac{1}{2}}$ and the Clebsch–Gordan coefficients for the angular momentum coupling $3/2 = 1/2 \oplus 1$
\begin{equation}
\begin{aligned}
\Pi^{s=3/2} & = \ket{\frac{3}{2}, \frac{3}{2}}\bra{1, 1; \frac{1}{2}, \frac{1}{2}}  + \ket{\frac{3}{2}, \frac{1}{2}} \left( \frac{1}{\sqrt{3}} \bra{1, 1; \frac{1}{2}, -\frac{1}{2}} \right. \\
+& \left. \sqrt{\frac{2}{3}} \bra{1, 0; \frac{1}{2}, \frac{1}{2}} \right) + \ket{\frac{3}{2}, -\frac{1}{2}} \left( \sqrt{\frac{2}{3}} \bra{1, 0; \frac{1}{2}, -\frac{1}{2}} \right. \\
+& \left. \frac{1}{\sqrt{3}} \bra{1, -1; \frac{1}{2}, \frac{1}{2}} \right) + \ket{\frac{3}{2}, -\frac{3}{2}}\bra{1, -1; \frac{1}{2}, -\frac{1}{2}},
\end{aligned}
\end{equation}
where the numbers in the kets and bras are abbreviations of the states $\ket{s, m^z}$. We obtain the following the MPS representations
\begin{equation}
\begin{aligned}
\ket{\Psi_\text{odd}}=\sum_{\{\mu_i\}} \text{Tr} & \left[A^{\mu_1}(B^{\mu_1}) A^{\mu_2}(B^{\mu_2}) \cdots A^{\mu_L}(B^{\mu_L})\right]  \\
& \ket{\mu_1 \mu_2 \cdots \mu_L},    
\end{aligned}
\end{equation}
where each tensor on the site $i$ could take the $A$ or $B$ tensor below, corresponding to the freely-rotating spin-$1/2$ taking $\ket{s=\frac{1}{2}, m^x = \frac{1}{2}}_i$ or $\ket{s=\frac{1}{2}, m^x = -\frac{1}{2}}_i$.
\begin{equation}
\begin{array}{ll}
A^{[\frac{3}{2}]}=\left(\begin{array}{cc}
0 & \sqrt{3} \\
0 & 0
\end{array}\right) 
& B^{[\frac{3}{2}]}=\left(\begin{array}{cc}
0 & \sqrt{3} \\
0 & 0
\end{array}\right) \\
A^{[\frac{1}{2}]}=\left(\begin{array}{cc} 
-1 & 1 \\
0 & 1
\end{array}\right) 
& B^{[\frac{1}{2}]}=\left(\begin{array}{cc}
-1 & -1 \\
0 & 1
\end{array}\right) \\
A^{[-\frac{1}{2}]}=\left(\begin{array}{cc} 
-1 & 0 \\
-1 & 1
\end{array}\right) 
& B^{[-\frac{1}{2}]}=\left(\begin{array}{cc}
1 & 0 \\
-1 & -1
\end{array}\right)\\
A^{[-\frac{3}{2}]}=\left(\begin{array}{ll}
0 & 0 \\
-\sqrt{3} & 0
\end{array}\right) 
& B^{[-\frac{3}{2}]}=\left(\begin{array}{cc}
0 & 0 \\
\sqrt{3} & 0
\end{array}\right).
\end{array}
\end{equation}

We mention that these matrices are related to those found in the DMRG-S calculations by appropriate MPS gauge transformations~\cite{Orus2014Practical,Cirac2021Matrix}.
In the following we prove that $\ket{\Psi_\text{odd}} \otimes \left(\bigotimes_{ i \in \text{even} } \ket{s=\frac{3}{2}, m^z=-\frac{3}{2}}_i \right) $ is a scarred eigenstate of 1D spin-$3/2$ PXP model with energy $E = L/2 - N_B $, where $N_B$ is the number of $B$ tensors appearing in $\ket{\Psi_\text{odd}}$. 

Since $A^{[-\frac{3}{2}]}A^{[-\frac{3}{2}]} = A^{[-\frac{3}{2}]}B^{[-\frac{3}{2}]} = B^{[-\frac{3}{2}]}A^{[-\frac{3}{2}]}  = B^{[-\frac{3}{2}]}B^{[-\frac{3}{2}]}=0 $, we have $P_i P_{i + 2} \ket{\Psi_\text{odd}} = 0 \  \forall \ \text{odd } i$.  In order to show that $\left( \sum_{ i \in \text{odd} }  S^x_i  \right) \ket{\Psi_\text{odd}} = (L/2 - N_B) \ket{\Psi_\text{odd}}$, we notice that the transformed $A(B)$ tensors by the spin-$3/2$ $S^x_i$ operator 
\begin{equation}
F^{\mu_i}_{A(B)} = \sum_{\mu_i'=-\frac{3}{2}}^{\frac{3}{2}} (S^x_i)^{\mu_i}_{\mu_i'} A^{\mu_i'}(B^{\mu_i'} )
\end{equation}
satisfy the condition
\begin{equation}
F^{\mu_i}_A = (\sigma^x A^{\mu_i}- A^{\mu_i} \sigma^x + A^{\mu_i} )/ 2,
\end{equation}
\begin{equation}
F^{\mu_i}_B = (\sigma^x B^{\mu_i}- B^{\mu_i} \sigma^x - B^{\mu_i}) / 2.
\end{equation}
Hence, by telescoping summing the series as in Sec.~\ref{sec:tilted_FH}, we obtain the desired eigenenergy $E = L/2 - N_B $.

\section{Symmetries of the exact scars in the 1D tilted Fermi-Hubbard model}
\label{sec:symm_analysis}

As shown by Ref.~\cite{Desaules2021Proposal}, in the Hilbert subspace dynamically connected to the scar states Eq.~\eqref{Eq:24_42_scar}, the effective Hamiltonian Eq.~\eqref{Eq:H_eff_spin} hosts three commuting symmetry operators $(\mathbb{Z}_1, \mathbb{Z}_2, \mathbb{S}^2)$.
The first one is the product of the doublon parity operator and the spin inversion operator
\begin{equation}
\mathbb{Z}_1=\prod_{j=1}^L (-1)^{P_{j,\downarrow} P_{j,\uparrow}} \left(S_{j, \uparrow}^+ S_{j, \downarrow}^- + S_{j, \downarrow}^+ S_{j, \uparrow}^- + \frac{1+ S^z_{j, \uparrow} S^z_{j, \downarrow}}{2} \right).
\end{equation}
The second symmetry is the joint action of the spatial inversion operator and the particle-hole conjugation
\begin{equation}
\mathbb{Z}_2=\prod_{j=1}^{L / 2} \prod_{\sigma = \uparrow, \downarrow} \left(S_{j, \sigma}^+ S_{\bar{j}, \sigma}^+ +  S_{j, \sigma}^- S_{\bar{j}, \sigma}^- + \frac{ 1- S^z_{j,\sigma} S^z_{\bar{j}, \sigma} }{2} \right),
\end{equation}
where $\bar{j} = L-j+1$ denotes the spatial inversion of the site $j$.
The third symmetry corresponds to the conservation of the total angular momentum $\mathbb{S}^2$ \cite{essler2005one}, defined as 
\begin{equation}
\begin{aligned}
&\mathbb{S}^2 = (\sum _j \mathbb{S}^x_j)^2 + (\sum _j \mathbb{S}^y_j)^2 + (\sum _j \mathbb{S}^z_j)^2, \\
&\mathbb{S}^\alpha_j= \frac{1}{2} \sum_{\beta,\gamma=\downarrow,\uparrow}  S_{j,\beta}^+ \left(\sigma^{\alpha} \right)_{\beta \gamma} S_{j,\gamma}^-.
\end{aligned}  
\end{equation}
$\sigma^{\alpha}$ ($\alpha=x,y,z$) are standard $2\times 2$ Pauli matrices.
The eigenvalues of $\mathbb{Z}_1$ and $\mathbb{Z}_2$ operators take $Z_{1(2)}=\pm 1$. The eigenvalues of the total angular momentum $\mathbb{S}^2$ could take $\{S(S+1)\}_{S=0}^{L/2}$.

The transformation of $\mathbb{Z}_1$ exchanges $B^{[\uparrow]}$ ($C^{[\uparrow]}$) with $B^{[\downarrow]}$ ($C^{[\downarrow]}$), and multiplies $B^{[\updownarrow]}$ ($C^{[\updownarrow]}$) with a $-1$ factor. That leads to the transformation on the block representation 
\begin{equation}
\tilde{A}^{[\uparrow,\downarrow]} = A^{[\downarrow,\uparrow]},\quad \tilde{A}^{[\downarrow,\uparrow]} = A^{[\uparrow,\downarrow]}, \quad \tilde{A}^{[\updownarrow,0]} = -A^{[\updownarrow,0]},
\end{equation}
which is equivalent to the MPS gauge transformation of $\sigma^x$. Specifically, 
\begin{equation}
\begin{aligned}
& v_\pm \tilde{A}^{d_1} \cdots \tilde{A}^{d_{\frac{L}{2}}} v_\mp^T \\
=& (v_\pm \sigma^x) (\sigma^x \tilde{A}^{d_1} \sigma^x)  \cdots  (\sigma^x \tilde{A}^{d_{\frac{L}{2}}} \sigma^x) (\sigma^x v_\mp^T) \\
=& - v_\pm A^{d_1}  \cdots A^{d_{\frac{L}{2}}} v_\mp^T.
\end{aligned}
\end{equation}
Thus, $\mathbb{Z}_1 \left|\Lambda_{\pm, \mp}\right\rangle =- \left|\Lambda_{\pm, \mp}\right\rangle$. Similarly, $\mathbb{Z}_1 \left|\Lambda_{\pm, \pm}\right\rangle = \left|\Lambda_{\pm,\pm}\right\rangle$.
 
The transformation of $\mathbb{Z}_2$ is the joint action of the spatial inversion $\mathcal{I}: j \rightarrow L-j+1$  and the particle-hole conjugation $\mathcal {C}$: $0 \rightarrow \updownarrow$, $\updownarrow \rightarrow 0$, $\downarrow \rightarrow \uparrow$ and $\uparrow \rightarrow \downarrow$, which transforms the $B$ and $C$ matrices Eq.~\eqref{Eq:B_C_matrix} as
\begin{equation}
\begin{aligned}
&B_\mathcal{CI}^{[0]} =  (C^{[\updownarrow]})^T ,\ B_\mathcal{CI}^{[\updownarrow]} =  (C^{[0]})^T, \ B_\mathcal{CI}^{[\uparrow]} = (C^{[\downarrow]})^T, \\ &B_\mathcal{CI}^{[\downarrow]} = (C^{[\uparrow]})^T, \ C_\mathcal{CI}^{[0]} = (B^{[\updownarrow]})^T , \ C_\mathcal{CI}^{[\updownarrow]} = (B^{[0]})^T, \\ &C_\mathcal{CI}^{[\uparrow]} = (B^{[\downarrow]})^T,\ C_\mathcal{CI}^{[\downarrow]} = (B^{[\uparrow]})^T.
\end{aligned}
\end{equation}
Then the transformation of the block representation is given by
\begin{equation}
A_\mathcal{CI}^{[\uparrow,\downarrow]} = (A^{[\uparrow,\downarrow]})^T,  A_\mathcal{CI}^{[\downarrow,\uparrow]} = (A^{[\downarrow,\uparrow]})^T,  
 A_\mathcal{CI}^{[\updownarrow,0]} = (A^{[\updownarrow,0]})^T.
\end{equation}

We can apply the MPS gauge transformation of $i\sigma^y$, specifically, 
\begin{equation}
\begin{aligned}
& v_\pm \tilde{A}^{d_1}  \cdots \tilde{A}^{d_{\frac{L}{2}}} v_\pm^T \\
=& (v_\pm i\sigma^y) (-i\sigma^y \tilde{A}^{d_1} i\sigma^y)  \cdots  (-i\sigma^y \tilde{A}^{d_{\frac{L}{2}}} i\sigma^y) (-i\sigma^y v_\pm^T) \\
=& (-1)^{L/2} v_\mp A^{d_1}  \cdots A^{d_{\frac{L}{2}}} v_\mp^T. 
\end{aligned}
\end{equation}
Thus, $\mathbb{Z}_2 \left|\Lambda_{\pm, \pm}\right\rangle =(-1)^{L/2} \left|\Lambda_{\mp, \mp}\right\rangle$. Similarly,
$\mathbb{Z}_2 \left|\Lambda_{\pm, \mp}\right\rangle =(-1)^{L/2+1} \left|\Lambda_{\pm, \mp}\right\rangle$.

\bibliographystyle{apsrev4-1-title}
\bibliography{DengQAIGroup,QMBS}

\end{document}